\documentclass[preprint,12pt]{elsart}
\usepackage{graphicx,amssymb,amsmath,times}
\usepackage{setspace}
\journal{Astroparticle Physics}

\begin{document}
\begin{frontmatter}
\title{Modelling the broadband emission from the white dwarf binary system AR Scorpii}
\author[1,2]{K K Singh\corauthref{cor}},
\corauth[cor]{Corresponding author.}
\ead{kksastro@barc.gov.in}
	\author[1]{P J Meintjes}, \author[1]{Q Kaplan}, \author[1]{F A Ramamonjisoa}, \author[2]{S Sahayanathan}
\address[1]{Physics Department, University of the Free State, Bloemfontein - 9300, South Africa}
\address[2]{Astrophysical Sciences Division, Bhabha Atomic Research Centre, Mumbai- 400 085, India} 

\begin{abstract}
AR Scorpii is a compact binary system which consists of a magnetic white dwarf and an M-type main sequence  
cool star. This binary system was discovered as a source of pulsating radiation in radio, infrared, optical, 
ultraviolet and X-ray wavebands. In this work, we have analyzed the $\gamma$-ray data in the energy range 100 MeV 
to 500 GeV from the \emph{Fermi}-Large Area Telescope (LAT) observations for the period August 4, 2008 to 
March 31, 2019. The $\gamma$-ray emission from AR Scorpii over the last decade is not statistically significant 
and therefore 2$\sigma$ upper limit on the integral flux above 100 MeV has been estimated. 
We reproduce the non-thermal broadband spectral energy distribution of AR Scorpii using an emission model having 
two synchrotron components due to the relativistic electrons in very high magnetic fields. The first component 
(Synchrotron-1) broadly describes the emissions at radio to high energy X-rays through the synchrotron radiation 
originating from a spherical region of radius $\sim$ 1.8$\times$10$^{10}$ cm and a magnetic field strength 
of $\sim$ 10$^3$ Gauss. 
The second component (Synchrotron-2) which reproduces the X-ray emission at lower energies and predicts the $\gamma$-ray 
emission, originates from another spherical region with radius $\sim$ 1.4$\times$10$^{10}$ cm and a magnetic 
field strength of $\sim$ 10$^6$ Gauss. The relativistic electron populations in both the emission regions are 
described by a smooth broken power law energy distribution. The $\gamma$-ray emission predicted by the 
Synchrotron-2 model is below the broadband sensitivity of the \emph{Fermi}-LAT and is also consistent with 
the 95$\%$ confidence level upper limit on the integral flux above 100 MeV derived from more than 10 years of 
observations. According to our model, the binary system AR Scorpii could be a $\gamma$-ray source, although its 
emission level must be below the current detection limit of the \emph{Fermi}-LAT.
\end{abstract}

\end{frontmatter}

\section{Introduction}  
White dwarfs (WDs) are formed as the end products of the evolution of intermediate main sequence stars in the cores of 
red giant stars [1,2]. They are considered as compact objects (electron-degenerate stellar configurations) with an 
average mass of $\sim$ 0.6M$_\odot$ and radius of the order of 10$^8$ cm ($\sim$ Earth-size). WDs have therefore a 
very high average density of 10$^6$ g cm$^{-3}$ and a large surface gravity. They are found as isolated objects or 
companions in various astrophysical binary systems with compact objects (black hole, neutron star, or  white dwarf), 
main sequence stars, giant stars or ordinary stars. Isolated WDs are observed with very high surface magnetic fields 
in the range 10$^3$ - 10$^9$ Gauss [3]. They are referred to as \emph{magnetic white dwarfs} (MWDs) and emit at 
ultraviolet and near-infrared wavelengths. The mass of MWDs ($\sim$ 0.8M$_\odot$) is more than the mass of 
non-magnetic or weakly-magnetic WDs ($\sim$ 0.5M$_\odot$) [3]. MWDs also exist in binary systems where they strip or 
accrete material from a nearby companion star. For main sequence companion stars, the accretion mainly 
occurs through Roche-lobe overflow whereas it is driven by the stellar wind in the case of red giant stars [4]. 
The resulting mass transfer from the main sequence companion stars to the MWDs produces atomic lines and X-ray emissions 
from the MWD binary systems [5,6]. In a typical WD binary system, the accreted matter heats up and forms a shock wave 
before settling down on the surface of the WD. Under the assumption of a dipole magnetic field, a MWD would behave like a 
radio pulsar [7]. The radio pulsars like Crab, characterized as the fast-rotating magnetized neutron stars, are assumed 
to be powered by the spin-down energy loss due to the rotation or spinning of neutron star [8]. The spin-down luminosity 
or rotational energy loss ($L_{down}$) of the pulsar or MWD is given by [9]
\begin{equation}
	L_{down}~=~ \frac{8 \pi^2 M R^2 \dot P}{5 P^3} 
\end{equation}
where $M$ and $R$ are the mass and radius of the WD respectively which are used to define the moment of inertia as 
$I = \frac{2}{5}M R^2$. $P$ and $\dot P$ are the spin period and its time derivative respectively. The magnetic spin-down 
luminosity ($L_{mag}$) for a magnetic dipole approximation of the MWD and assuming a wind outflow from the magnetosphere, 
can be approximated as [10]
\begin{equation}
	L_{mag}~=~\frac{16 \pi^4 B_s^2 R^6}{6 c^3 P^4}
\end{equation}	
where $B_s$ is the strength of the surface magnetic field at the polar cap and $c$ is the speed of light in vacuum. 
If the magnetic dipole losses of the WD dominate the luminosity of a binary system, the strength of the surface magnetic 
field of the MWD at the polar cap is related to the different parameters of the pulsar as [10]   
\begin{equation}
	B_s~=~\sqrt{\frac{3 M c^3 P \dot P}{5 \pi^2 R^4}}
\end{equation}	
and the light cylinder radius ($R_{LC}$) of the spinning MWD is defined by 
\begin{equation}\label{eqn:Rlc}
	R_{LC}~=~\frac{c P}{2 \pi}
\end{equation}
The magnetic field strength at the light cylinder ($B_{LC}$) is approximated as 
\begin{equation}
	B_{LC}~=~4 \pi^2 \sqrt{\frac{3 M R^2 \dot P}{5 c^3 P^5}}
\end{equation}	
Therefore, a precise measurement of $P$ and $\dot P$ is very important to derive the energetics of a pulsar or 
spinning MWD. $B_{LC}$ plays a very crucial role for the $\gamma$-ray emission from these sources due to the 
conversion of a significant fraction of the spin-down power ($L_{down}$) to $\gamma$-ray luminosity.
\par
The binary systems comprising the MWDs and main sequence stars are divided into two classes namely polar and intermediate 
polar depending on the strength of the surface magnetic field of the WD. For polar type, the surface magnetic field ($B_s$) 
is higher than 10$^7$ Gauss and accretion from the companion star is completely confined through the magnetic field of 
the WD [11]. In the case of intermediate polar class, $B_s$ is in the range 10$^3$ - 10$^6$ Gauss and the accretion process 
is partially channeled through the surface magnetic field [12]. Due to the stronger surface magnetic fields, the rotation of 
the two stars is synchronised in the polar binary systems. Whereas, in the intermediate polar systems, the rotation of the WD 
is not synchronised with the companion star and the spin period of the WD is shorter than the orbital period. The 
formation of an accretion disk around the WD is prevented in the polar binary systems because of the synchronised rotation 
of the two stars and no angular momentum of the accreting matter with respect to the WD. But, an accretion disk can be 
formed in the intermediate polar systems and the transfer of angular momentum from the accreting matter may spin up the WD [13]. 
The accretion flows in polars and intermediate polars lead to the formation of funnels and curtains respectively. The differential 
rotation of gaseous matter leaking out from the companion star along a circular Keplerian orbit leads to a viscous rubbing of 
fluid elements at varying distances and causes the accretion disk to heat up. A sufficiently hot accretion disk emits 
radiation from optical, ultraviolet to X-rays. The surface magnetic field in the intermediate polar disrupts the inner 
part of the accretion disk and the accreting matter along the magnetic field lines [13]. This leads to the formation of 
one or more accretion columns near the magnetic poles which emit non-thermal polarized radiation from radio to X-rays. 
The variations in the non-thermal emissions from the intermediate polar binary systems are strongly modulated on the 
spin period ($P$) of the MWD and its harmonics [14]. However, the exact physical process involved in the non-thermal broadband 
emission from the WD binary systems has not been clearly understood and is being widely debated. In this work, we model the 
time averaged broadband spectral energy distribution (SED) of an intermediate polar AR Scorpii using observations spanned 
over last two decades. In Section 2, we present a brief description  of the important observational features so far of 
AR Scorpii. The details of the broadband data set used in the present work are given in Section 3. In Section 4, we discuss 
the results from the multi-wavelength emission from the binary system AR Scorpii. Finally, we conclude the study in Section 5.
 
\section{AR Scorpii}
AR Scorpii is an intermediate polar type WD binary system at a distance of $d \sim$ 110 pc (3.4$\times$10$^{20}$ cm) 
from the Earth and is located in the ecliptic plane near to the Galactic centre [14]. It consists of an MWD and an 
M-type main sequence companion star with a binary separation of $ a \sim$ 7.6$\times$10$^{10}$ cm and both move 
around the common centre of mass. The mass and radius of the MWD in AR Scorpii are $M = 0.8M_\odot$ and 
$R = 7 \times 10^8$ cm respectively and the surface magnetic field $B_s \sim 10^8$ Gauss [14]. The companion M-star 
has a mass of $m \sim 0.3M_\odot$ and a radius of $r \sim 2.7\times 10^{10}$ cm. In early observations during 1970-71, 
AR Scorpii was classified as a periodic variable star ($\delta-$Scuti) and recently it was discovered as a close MWD binary 
system with an orbital period of $P_o \sim$ 3.56 hours and a spin period of the MWD $P \sim$ 1.95 minutes (117 seconds) at 
radio and optical wavelengths [14]. The measured time derivative of the spin period of the MWD in AR Scorpii was 
$\dot P \sim 3.9\times10^{-13}$ seconds/second. This gives an estimate of the age of the MWD as 
$t=\frac{P}{2 \dot P} \sim 4.5\times10^6$ years. The photometry of AR Scorpii taken over seven years suggests that the 
R-band optical magnitude changes from 16.9 (faintest) to 13.6 (brightest) on a time period of $\sim$ 3.56 hours. The amplitude 
spectra corresponding to the ultraviolet, optical, infrared and radio (9 GHz) fluxes show signals having two components, 
which are identified as the spin frequency ($\nu_s = P^{-1}$) and beat frequency ($\nu_b = \nu_s - \nu_o$, 
where $\nu_o = P_o^{-1}$ is the orbital frequency). The beat component of the signal is stronger and therefore it defines the 
dominant pulsation at a beat period of $P_b =$ 1.97 minutes and its harmonics [14]. The observations of AR Scorpii 
over a wide wavelength range from radio to ultraviolet show strong double-humped pulsations with the pulse fractions between 
10$\%$ to 95$\%$ [14]. The optical spectra of M-type cool main 
sequence star with absorption lines suggest that the radial velocity of the companion star in AR Scorpii varies sinusoidally 
with a time period of $\sim$ 3.56 hours and an amplitude of $\sim$ 290 km s$^{-1}$ [14]. The MWD in AR Scorpii is not visible 
in the optical spectra. For a spin period $P \sim 117$ seconds, the light cylinder radius of the MWD $R_{LC} \approx 5.6\times10^{11}$ cm 
(from Equation \ref{eqn:Rlc}) which is about seven times longer than the binary separation. This implies that the companion star 
in AR Scorpii is situated well inside the magnetosphere of the MWD. The MWD in AR Scorpii is spinning down on a 
characteristic timescale of $\sim$ 10$^7$ years [14]. The X-ray emission from AR Scorpii is significantly modulated over the spin 
period of the MWD with a pulse fraction of $\sim$ 14$\%$ [15]. The maximum or minimum X-ray intensity is located at the 
superior or inferior conjunction of the M-star orbit. An evidence of a power law spectrum is found in the pulsed component of 
the X-ray emission [15]. The observed optical emission from AR Scorpii modulating with the beat frequency of the binary system 
is highly polarized with a degree of linear polarization up to 40$\%$ and a low level of circular polarization of less than 
5$\%$ [16,17]. The degree of linear polarization evolves with the orbital phase and varies on both the beat frequency of the 
binary system and spin frequency of the MWD. The observed characteristics of the degree of linear polarization are similar to 
that of the well-known spin-powered Crab pulsar. The radio observations at 8.5 GHz reveal the compactness of the emission region 
and confirm the identity of AR Scorpii as a point Galactic source [18]. The high resolution interferometric radio imaging of 
AR Scorpii indicates a strong modulation of the radio flux on the orbital and beat periods and the radio emission exhibits 
a weak linear polarization and very strong circular polarization [19]. High temporal resolution spectroscopy of this MWD binary 
system shows a complex structure of the $H\alpha$ emission lines as observed from some polars in the quiescent state [20]. 
The aforementioned measurements/results suggest a non-thermal origin of the broadband radiation from AR Scorpii 
through the synchrotron emission process.

\section{Broadband Data Set}
In this work, we use archival results from the radio, infrared, ultraviolet, optical and X-ray observations of AR Scorpii 
available so far in the literature [14,15]. We have also analyzed data from the \emph{Fermi}-Large Area Telescope 
(LAT) observations for a period of more than 10 years to search for $\gamma$-ray emission from AR Scorpii. 
A brief description of the broadband archival data as well as a detailed analysis of the \emph{Fermi}-LAT data 
are given below.

\subsection{Archival Data}
We have obtained the spectral archival points at radio, infrared, ultraviolet, optical and X-ray energies 
from Figure 4 of Marsh et al. (2016) [14] and Figure 10 of Takata et al. (2018) [15]. These flux measurements in different wavebands 
include observations from a number of ground and space-based instruments during the period 2006-2016. The X-ray 
flux points correspond to the time averaged spectrum in the energy range 0.15-12 keV from the \emph{XMM-Newton} 
observations of AR Scorpii on September 19, 2016. The \emph{XMM-Newton} observation covers more than two orbits 
of AR Scorpii and the X-ray emission shows a large variation over the orbit. A detailed description of 
the \emph{XMM-Newton} data can be found in [15].

\subsection{\emph{Fermi}-LAT Data}
The Large Area Telescope (LAT) onboard the \emph{Fermi} satellite is a pair-conversion detector to measure the energy, 
arrival time, and direction of the $\gamma$-ray photons in the energy range 30 MeV to more than 1 TeV [21]. It also 
detects the intense cosmic ray background of the charged particles and radiation trapped at the orbit of the satellite. 
For $\gamma$-ray photons at 1 GeV energy, the \emph{Fermi}-LAT has an effective collection area of $\sim$ 1 m$^2$ at normal 
incidence, a wide field of view of 2.4 steradians (20$\%$ of the sky), and an angular resolution of 0.8$^\circ$. 
At lower energies, the photon angular resolution is poor ($\sim$ 5$^\circ$ at 0.1 GeV) and it improves with the 
increase in energy ($\sim$ 0.1$^\circ$ above 10 GeV). Therefore, the angular resolution of the \emph{Fermi}-LAT can be 
approximated as a power law function of energy (E$^{-0.8}$, where E is the energy of the $\gamma$-ray photon) in its 
operational energy range. The whole sky coverage is provided by the \emph{Fermi}-LAT every three hours in survey mode 
of its operation.
\par
In this work, we have extracted the Pass 8 Release 3 (P8R3) data for the position of AR Scorpii 
(RA: 16$^h$21$^m$47.28$^s$, Dec: -22$^\circ$53$^\prime$10.39$^{\prime\prime}$) collected through the 
last decade by the \emph{Fermi}-LAT during August 4, 2008 to March 31, 2019 (MJD 54682-58573) from 
the \emph{Fermi} Science Support Centre\footnote{https://fermi.gsfc.nasa.gov/cgi-bin/ssc/LAT/LATDataQuery.cgi}. 
The data is processed using standard \emph{Fermi} Science Tools software version 1.0.1 (Fermi 1.0.1). We have considered 
only SOURCE class events ($evclass=128$, $evtype=3$) corresponding to the $P8R3\_SOURCE\_V6$ instrument response function 
detected within a circular region of interest (ROI) of 10$^\circ$ radius centred on the position of AR Scorpii for 
the above period in the energy range 100 MeV to 500 GeV using the \emph{gtselect} tool. In addition, a maximum zenith 
angle cut of 90$^\circ$ is applied to avoid the albedo contamination from the Earth limb $\gamma$-rays while creating the good 
time intervals. We have used unbinned maximum-likelihood technique implemented in the \emph{gtlike} tool to further 
analyse the dataset. A background model is generated using all the sources reported in the fourth \emph{Fermi} gamma-ray 
catalog (4FGL) [22] and recent Galactic diffuse and extragalactic isotropic background emission templates. All the 
4FGL point sources present within the ROI are included in the model file for background subtraction keeping their 
spectral forms the same as defined in the 4FGL catalog [22]. Since the target source AR Scorpii is not reported in the 4FGL catalog, 
we have manually added this source in the model file with a spectral shape defined by a simple power law function. We use an iterative method 
to optimize the contribution of sources in the background model file. The associated spectral parameters of all the sources within the 
ROI including AR Scorpii are left free to vary during the unbinned likelihood fitting. The significance of $\gamma$-ray signal is 
estimated from the maximum-likelihood test statistic (TS: square-root of TS gives the statistical significance) as defined in [23]. 
After every fit, we remove the background sources from the model file corresponding to a value of TS$<$25 (statistical 
significance $\sim$ 5$\sigma$) and the process is repeated until we get all the background sources with TS$\ge$25 within the ROI 
in the model file except the target source AR Scorpii. We perform the final unbinned likelihood fitting by allowing the spectral parameters 
of AR Scorpii and other background sources within 5$^\circ$ from the center of ROI to vary and the spectral parameters of all other sources 
lying beyond 5$^\circ$ are freezed to the values obtained from the optimization. 

\begin{figure}
\begin{center}
\includegraphics[width=0.70\textwidth,angle=-90]{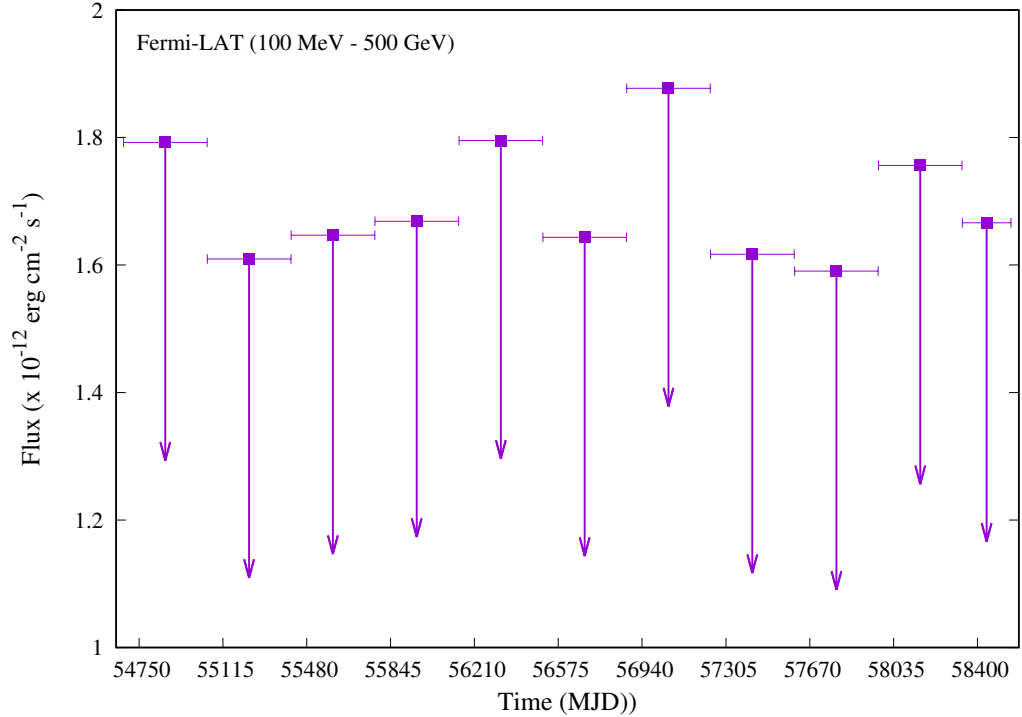}
	\caption{One year binned (except the last data point which is averaged over about eight months) $\gamma$-ray light curve of 
	the white dwarf binary system AR Scorpii from the \emph{Fermi}-LAT observations between August 4, 2008 and March 31, 2019 (MJD 54682-58573). 
	The downward arrows indicate 2$\sigma$ upper limit on the integral energy flux above 100 MeV.}
\label{fig:Fig1}
\end{center}
\end{figure}

\section{Results and Discussion}

\subsection{Search for $\gamma$-ray emission from AR Scorpii}
The Pass 8 dataset from the \emph{Fermi}-LAT observations provides a better measurement of the energy of 
$\gamma$-ray events over a wider energy range with their improved reconstruction and a significant increase in the 
effective area at lower energies. From the unbinned-likelihood fitting as described above, we observe that the 
\emph{Fermi}-LAT data collected in the direction of AR Scorpii for a period of more than ten years assuming a 
power law spectral index of -2.7 in the energy range of 100 MeV-500 GeV, exhibit a TS value of 8. 
This implies that the $\gamma$-ray emission from AR Scorpii integrated over a decade from the \emph{Fermi}-LAT 
observations is not statistically significant ($<$3$\sigma$) and the ratio of the error in the integral flux to the 
absolute value is more than 0.5. Therefore, we have estimated the 2$\sigma$ upper limit of 
2.27$\times$10$^{-12}$ erg~cm$^{-2}$~s$^{-1}$ on the integral energy flux above 100 MeV for the 
$\gamma$-ray emission from the binary system AR Scorpii. The low statistical significance of the $\gamma$-ray observation 
from AR Scorpii is consistent with the fact that this source is not reported in the latest 4FGL catalog which includes more than 5000 sources 
above 4$\sigma$ significance [22]. We have also generated the yearly light curve for AR Scorpii to observe any temporal fluctuation 
in the $\gamma$-ray emission from this source. The one year binned light curve of AR Scorpii in the energy range 100 MeV to 500 GeV 
is shown in Figure \ref{fig:Fig1}. All the flux points in the light curve (Figure \ref{fig:Fig1}) correspond to the 2$\sigma$ upper 
limit on the integral flux. This indicates that AR Scorpii is not a $\gamma$-ray source above the current detection 
sensitivity of the \emph{Fermi}-LAT. The Galactic diffuse emission is bright in the MeV-GeV energy regime 
and provides a dominant contribution to the background radiation near the Galactic plane. Due to a relatively broad angular resolution of 
the \emph{Fermi}-LAT at lower energies, the emission from a faint $\gamma$-ray source like AR Scorpii may be contaminated by the 
diffuse background or photons from the nearby sources. This can affect the search for a pulsed signal from the weak $\gamma$-ray sources. 
An event weighting technique can be used to search for a pulsation from a faint $\gamma$-ray source [24]. The search for 
pulsed $\gamma$-ray signal from the WD binary system AR Scorpii is beyond the scope of this work.

\begin{figure}
\begin{center}
\includegraphics[width=0.70\textwidth,angle=-90]{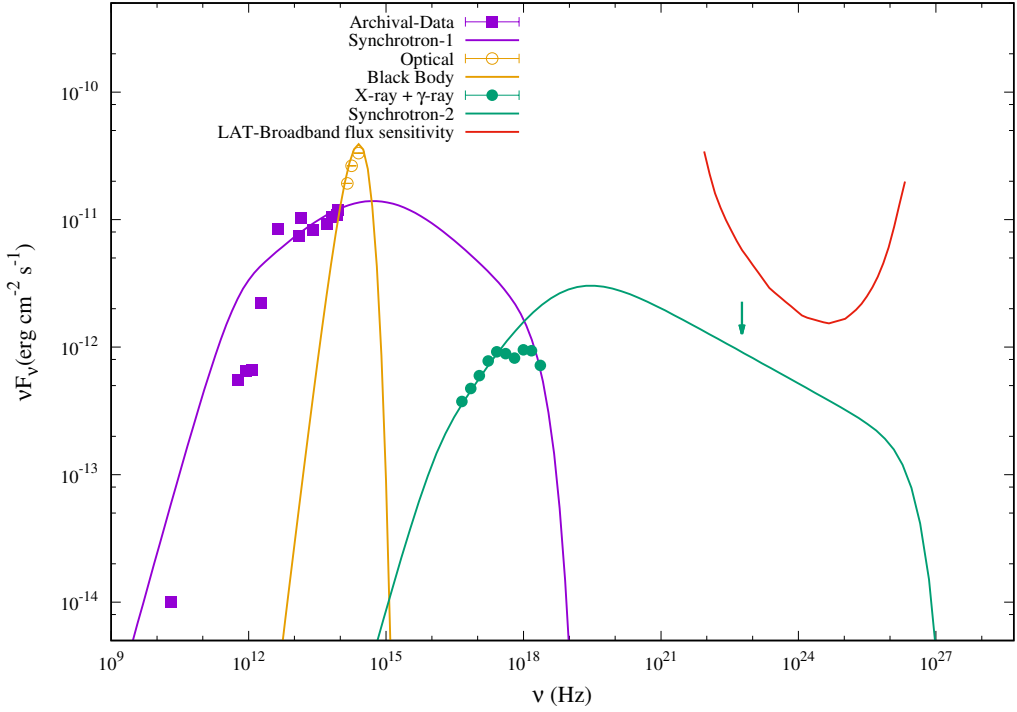}
	\caption{Broadband spectral energy distribution of AR Scorpii with the time averaged observed data points without 
	measurement errors and model curves. The curves with purple and green colours correspond to the Synchrotron-1 and 
	Synchrotron-2 components of the model for a time averaged non-thermal emission from the binary system AR Scorpii 
	respectively. The yellow curve is associated with the black body emission at a temperature of $T_b \sim$ 3100 K from 
	the companion M-type star of radius $r = R_b \sim 3\times10^{10}$ cm in the binary system. The data point with 
	downward arrow represent 2$\sigma$ upper limit on the integral flux above 100 MeV from the \emph{Fermi}-LAT observations 
	over a decade. The measured 2$\sigma$ upper limit and non-thermal emission predicted by the Synchrotron-2 component of 
	the model indicate that the $\gamma$-ray emission from the source is very weak and below the current flux sensitivity of 
	the \emph{Fermi}-LAT, depicted by the red curve (taken from the \emph{Fermi}-LAT Performance at the \emph{Fermi}
	Science Support Center).}
\label{fig:Fig2}
\end{center}
\end{figure}

\subsection{Broadband Spectral Energy Distribution Modelling}
We have employed a leptonic synchrotron model with two emission components to reproduce the time averaged broadband SED of the 
binary system AR Scorpii. We assume that the non-thermal radiation from AR Scorpii is produced by the population of 
relativistic electrons through the synchrotron emission. The electron energy distribution is described by a smooth broken-power 
law of the form
\begin{equation}\label{elec-spec}
N(\gamma)~=~K\left[\left(\frac{\gamma}{\gamma_b}\right)^p + \left(\frac{\gamma}{\gamma_b}\right)^q\right]^{-1}~;~~\gamma_{min} < \gamma < \gamma_{max}
\end{equation}	
where $\gamma=\frac{E}{m_e c^2}$ is the typical Lorentz factor of the electron with $m_e$ being the electron rest mass. $K$ is the 
normalization coefficient, $p$ and $q$ are the low and high energy electron spectral indices before and after the break respectively. 
$\gamma_b$, $\gamma_{min}$ and $\gamma_{max}$ are the Lorentz factors corresponding to the break, minimum and maximum energies of the 
electrons respectively. The synchrotron emissivity due to an isotropic electron distribution in a tangled magnetic field ($B$) 
is calculated as 
\begin{equation}
	j_{syn}(\nu)~=~\frac{1}{4\pi} \int_{\gamma_{min}}^{\gamma_{max}} N(\gamma)P(\nu,\gamma)d\gamma
\end{equation}
where $\nu$ is the frequency of the synchrotron photon and $P(\nu,\gamma)$ is the single electron synchrotron emissivity averaged 
over an isotropic distribution of pitch angles [25,26]. $P(\nu,\gamma)$ can be expressed as 
\begin{equation}
	P(\nu,\gamma)~=~\frac{\sqrt{3} \pi e^3 B}{4 m_e c^2} F\left(\frac{\nu}{\nu_c}\right)
\end{equation}
with the critical frequency,
\begin{equation}\label{nuc}
	\nu_c~=~\frac{3 \gamma^2 e B}{16 m_e c}
\end{equation}	
where $e$ is the charge of the electron, and $F\left(\frac{\nu}{\nu_c}\right)$ is the synchrotron power function, which 
can be approximately given by [26]
\begin{equation}
	F\left(\frac{\nu}{\nu_c}\right)~\approx~1.8 \left(\frac{\nu}{\nu_c}\right)^{1/3} \rm e^{-\frac{\nu}{\nu_c}} 
\end{equation}
The synchrotron power function peaks at $\frac{\nu}{\nu_c}$ = 0.29. This indicates that the peak frequency ($\nu_p$) in 
the synchrotron spectrum is about 29$\%$ of the characteristic frequency ($\nu_c$) defined by Equation \ref{nuc}. 
The synchrotron flux received by the observer from the source is given by 
\begin{equation}
	F_{obs}(\nu)~=~\frac{V}{d^2}j_{syn}(\nu)
\end{equation}
where $d$ is the luminosity distance ($\sim$ distance) of the source and $V$ is the volume of the emission region. 
Assuming that the emission region is a homogeneous sphere of radius $R_{em}$ filled with the relativistic electrons 
defined by Equation \ref{elec-spec} and permeated in a tangled magnetic field of strength $B$, we have reproduced 
the time averaged broadband SED of AR Scorpii. The model curves along with the data points in different wavebands are 
shown in Figure \ref{fig:Fig2}. We observe that the time averaged broadband emission from the binary system AR Scorpii 
can be broadly reproduced by two non-thermal components referred to as \emph{Synchrotron-1} and \emph{Synchrotron-2}. 
The multi-wavelength data points in Figure \ref{fig:Fig2} have been used without taking into account the measurement errors 
from different instruments and are reprocduced through the $\chi$-by-eye fitting process. It is evident from Figure \ref{fig:Fig2} 
that the  Synchrotron-1 component of the model describes the emission in the low energy band (radio-IR-Optical-UV), whereas 
the Synchrotron-2 component predicts soft X-ray and $\gamma$-ray emissions from the binary system AR Scorpii. The high energy end 
of the X-ray emission (hard X-rays) in the energy range 0.15-12 keV is reproduced by the tail of the synchrotron spectrum from 
the Synchrotron-1 component of the model. A comparison of the $\gamma$-ray emission level predicted by the Synchrotron-2 model 
with the broadband flux sensitivity of the \emph{Fermi}-LAT in Figure \ref{fig:Fig2} suggests that the $\gamma$-ray flux above 100 MeV 
expected from AR Scorpii is below the current detection sensitivity of the LAT as well as the 2$\sigma$ upper limit on the integral 
flux derived from more than 10 years of observations. The best fit model parameters derived from the non-thermal broadband SED 
modelling of AR Scorpii are given in Table \ref{tab:sed-par}. We note that the size of the emission region ($R_{em}$) for the 
Synchrotron-1 and Synchrotron-2 components is much smaller than the light cylinder radius ($R_{LC}$) of the MWD and/or the 
binary separation ($a$) in AR Scorpii. This implies that the non-thermal emission zones are situated well inside the light cylinder. 
The magnetic field strength in the X-ray and $\gamma$-ray emission regions (Synchrotron-2) is much higher than that in the low energy emission 
zone (Synchrotron-1). However, the strength of the magnetic field in both the emission regions is significantly less than the 
surface magnetic field ($B_s$) of the MWD in AR Scorpii. The dipole magnetic field of the MWD at a radial distance $x$ is given by 
\begin{equation}
	B_x~=~B_s \left(\frac{R}{x}\right)^3
\end{equation}	
Using this relation, we get an estimate of the locations of the emission regions from the MWD as $x_1 = 3.1\times10^{10}$ cm and 
$x_2 = 3.1\times10^9$ cm corresponding to the Synchrotron-1 and Synchrotron-2 components respectively. This implies that 
the low energy emission zone (Synchrotron-1) is near to the M-type companion star ($x_1 \sim a \sim 10^{10}$ cm), whereas the high 
energy emission zone  (Synchrotron-2) is located between the MWD and M-type companion star ($x_2 < a$) in the binary system.
\par
The particle spectral indices $p$ and $q$ describing the smooth broken-power law energy distribution (Table \ref{tab:sed-par}) 
indicate an efficient acceleration of electrons to relativistic energies and synchrotron cooling in the binary system AR Scorpii. 
The energy density of relativistic electrons ($U_e$) in the emission region is given by 
\begin{equation}\label{ene-dens}
	U_e~=~m_e c^2 \int_{\gamma_{min}}^{\gamma_{max}} N(\gamma)d\gamma .
\end{equation}
From Table \ref{tab:sed-par}, we find that the energy density of the electrons in the high energy emission zone (Synchrotron-2) 
is about 5 times higher than that in the low energy emission region (Synchrotron-1). This indicates that the electrons 
associated with the Synchrotron-2 component are accelerated to higher relativistic energies ($\gamma_{max} \sim 10^6$) than those 
involved in the Synchrotron-1 component. Therefore, electrons near to the MWD experience relatively more acceleration as 
compared to the far away electrons. The values of $p$ and $q$ for both the emission components are very close to the condition 
of radiative cooling break $q = p+1$. The synchrotron cooling time scale for an electron is defined as [27]
\begin{equation}
	t_{sync}~=~\frac{9}{4} \frac{m_e^3 c^5}{e^4 B^2 \gamma}~=~7.7 \times 10^4 \left(\frac{10 G}{B}\right)^2 \left(\frac{10^2}{\gamma}\right)~~\rm{second}.
\end{equation}
If the maximum energy of an electron in the emission region is determined by the synchrotron cooling, the acceleration timescale should 
be the same as $t_{sync}$. Therefore, assuming $\gamma = \gamma_{max}$, the acceleration timescale of an electron in AR Scorpii can be 
expressed as  
\begin{equation}
	t_{acc}~=~7.7 \times 10^4 \left(\frac{10 G}{B}\right)^2 \left(\frac{10^2}{\gamma_{max}}\right)~~\rm{second}.
\end{equation}
Using the values of $B$ and $\gamma_{max}$ from Table \ref{tab:sed-par}, we find that $t_{acc}$ is very small and therefore, suggests a very 
fast and efficient acceleration of electrons in AR Scorpii. The exact acceleration process for the electrons in the WD binary systems 
remains unclear. However, several plausible scenarios have been proposed for AR Scorpii. In the first scenario [28], a magnetized 
plasma ejected from the polar cap of the MWD sweeps the stellar wind from the M-type companion star and interaction between the 
plasma and stellar wind leads to the formation of a bow shock. This bow shock propagates in the stellar wind and accelerates the 
electrons to relativistic energies in the wind to produce synchrotron radiation. In the second scenario [29], the magnetic interaction 
on the surface of the M-type companion star has three dissipation effects namely heating of the surface of M-star, an overflow from the M-star 
and acceleration of electrons to the relativistic energies with $\gamma =$ 50-100. The accelerated electrons move from the companion star 
to the surface of the MWD along the magnetic field lines and are trapped in the closed magnetic field lines region due to the magnetic 
mirror effect. The synchrotron radiation is produced by the trapped electrons at the magnetic mirror point. In the third scenario [30], 
the particles are accelerated through the magnetic reconnection or \emph{Fermi} acceleration process in a turbulent collision region 
produced by the interaction of the MWD magnetic field with a dense atmosphere of the M-type star. In a relatively different hybrid scenario [31], 
electrons and protons are accelerated to the relativistic energies close to the surface of the companion M-star as postulated in the third 
scenario and the broadband non-thermal emission is produced by the relativistic electrons and/or by the electron-positron pairs from the 
pion decay created by the interaction of relativistic protons with the matter in the atmosphere of the M-star. Under the framework of 
the bow shock acceleration, the electron energy distribution has been reasonably described by a broken-power law with $p \sim 2.4$ and 
$q = p+1$ constrained by the X-ray observations [28]. The measurements of the optical linear polarization from AR Scorpii support the bow shock 
acceleration of electrons [16]. Therefore, the electron energy distribution described by a smooth broken-power law in the present study can be 
attributed to the bow shock acceleration scenario.
\begin{table}
\caption{Best fit model parameters from the non-thermal spectral energy distribution  modelling of the binary system AR Scorpii.}
\begin{center}
\begin{tabular}{lcccc}
\hline
Parameter		   		&Symbol      	&Synchrotron-1		&Synchrotron-2\\
\hline
Size of emission region  			&$R_{em}$	&1.8$\times10^{10}$ cm  &1.4$\times10^{10}$ cm \\		
Magnetic field strength  			&$B$ 		&1.1$\times10^3$ G	&1.1$\times10^6$ G\\
Low energy index of particle ditribution 	&$p$		&2.3			&1.8\\
High energy index of particle ditribution 	&$q$          	&3.7 			&3.4\\
Break energy of particle distribution     	&$\gamma_b m_e c^2$	&200 MeV 	&715 MeV\\
Minimum Lorentz factor of particle distribution &$\gamma_{min}$ &10			&50\\
Maximum Lorentz factor of particle distribution &$\gamma_{max}$ &2$\times10^4$		&8$\times10^6$\\
Particle energy density                  	&$U_e$          &0.46 erg~cm$^{-3}$	&2.40 erg~cm$^{-3}$\\
\hline
\end{tabular}
\end{center}
\label{tab:sed-par}
\end{table}
\par
However, from Figure \ref{fig:Fig2}, we observe that a hump at the optical frequencies cannot be reproduced by 
the non-thermal synchrotron emission predicted by the Synchrotron-1 model. Therefore, we use a simple black body emission to 
model this optical emission from AR Scorpii. The spectral energy distribution for a thermal emission from a black body 
is estimated as  
\begin{equation}
	\nu F_\nu~=~8 \pi \rm h \left(\frac{R_b}{d}\right)^2 \frac{\nu^4}{c^2} \left(\rm e^{\rm h \nu/k T_b} - 1\right)^{-1}
\end{equation}
where $R_b$ and $T_b$ are the radius and temperature of the black body respectively. $h$ and $k$ are the Planck and Boltzmann constants 
respectively. From the best fit of the optical hump in Figure \ref{fig:Fig2}, we get $R_b = 3.2\times10^{10}$ cm and $T_b =$ 3100 K. These 
parameters are in good agreement with the values reported in the literature [14,15,28]. Hence, the black body can be associated with the 
companion M-type star in AR Scorpii. This supports the above finding that the Synchrotron-1 emission zone is near to the M-star in the 
binary system. Therefore, the thermal emission from the M-type star dominates over the synchrotron radiation at the optical frequencies. 
The dominant thermal emission from the M-type star can play a major role in the depolarization of the synchrotron radition at the 
optical wavelengths. Therefore, the thermal contribution of the M-type star should be taken into account while modelling the synchrotron 
polarization of the source. The intrinsic synchrotron polarization modelling will help to better understand the acceleration process in AR Scorpii.

\section{Conclusions}
We have used the broadband archival data from radio to X-ray observations to reproduce the time averaged spectral energy 
distribution of the binary system AR Scorpii under the framework of a leptonic emission model. We have also derived a 
2$\sigma$ upper limit on the $\gamma$-ray integral flux above 100 MeV by analysing the \emph{Fermi}-LAT data for a period of 
more than a dacade. The important findings of this study are as follows:
\begin{itemize}
	\item The synchrotron emission produced by the relativistic electrons described by a smooth broken-power law 
		energy distribution from two independent components located well inside the light cylinder of the MWD 
		contributes to the observed broadband SED of AR Scorpii. 

      \item  Low energy observations in the radio, infrared, optical and ultraviolet frequency bands are broadly reproduced 
	      by the non-thermal emission which originates from a spherical region near the M-type companion star. In this region, 
		we find that the strength of the magnetic field is $\sim$ 10$^3$ G and the electron energy distribution is decribed 
		by the spectral indices 2.3 and 3.7 before and after the break respectively with a break energy of $\sim$ 200 MeV.

     \item  Most of the X-ray flux points in the energy range 0.15-12 keV are reproduced by the synchrotron emission from a spherical 
	    zone with a magnetic field strength of $\sim$ 10$^6$ G and located at a distance of $\sim$ 10$^9$ cm from the MWD. 
		The relativistic electrons in this region also follow an energy distribution with spectral indices 1.8 and 3.4 
            before and after the break respectively and having a break at $\sim$ 715 MeV. The flux points at the high energy end 
		of the X-ray spectra are described by the tail of the synchrotron radiation which is dominant at the low energy 
		component of the broadband SED of AR Scorpii.

	\item The measured optical emission from AR Scorpii indicates a hump in the broadband SED and cannot be modelled by any 
		of the non-thermal emission components. Instead, the optical hump is well reproduced by a black body spectrum 
		associated with the M-type companion star in the binary system.
  
     \item  The high energy $\gamma$-ray emission level predicted by the SED modelling of the source is below the 2$\sigma$ upper 
	     limit on the integral energy flux above 100 MeV as well as the current broadband flux sensitivity of the \emph{Fermi}-LAT. 
	     This suggests that if the binary system AR Scorpii emits gamma rays, it would be a weak Galactic source, but we cannot 
		make an implicit assumption that it is. However, it can be a potential target for the upcoming ground-based 
		$\gamma$-ray observatory namely the Cherenkov Telescope Array (CTA) with a low energy threshold and better sensitivity.  	
\end{itemize}
\section*{Acknowledgements}
Authors thank the anonymous reviewer for the important and valuable suggestions to improve the manuscript. 
We acknowledge the use of public data obtained through the Fermi Science Support Center (FSSC) provided by NASA, and other 
archival data from various observations in this study.

\end{document}